\documentstyle{article}
\newcommand{\be}{\begin{equation}}
\newcommand{\ee}{\end{equation}}
\def\bea{\begin{eqnarray}}
\def\eea{\end{eqnarray}}
\def\al{\alpha}
\def\ba{\beta}
\def\ga{\gamma}
\def\da{\delta}
\def\la{\lambda}
\def\sa{\sigma}
\def\om{\omega _{k}}

\def\P{P^{\al_{1}\ba_{1}...\al_{n}\ba_{n}\al_{n+1}}(T)}
\def\lb{\label}
\newcommand{\T}{T^{\al\ba\ga\da}}
\newcommand{\bb}{\bibitem}
\newcommand{\f}{\frac{1}{2}}
\newcommand{\k}{\frac{1}{4}}
\newcommand{\p}{\phi}
\newtheorem{th}{Theorem}
\begin{document}
\title{Superenergy tensors for a massive scalar field}
\author{Pierre Teyssandier\\
\small Laboratoire de Gravitation et Cosmologie Relativistes,
CNRS/ESA 7065,\\
\small Universit\'e Pierre et Marie Curie, Tour 22/12, BP 142,\\
\small 4 Place Jussieu, F-75252 Paris Cedex 05, France}
\date{\small May 5, 1999}
\maketitle

\begin{abstract}
We define a general class of superenergy tensors of even rank 
$2(n+1)$ for a real massive scalar field propagating in Minkowski spacetime. 
In the case where $n=1$, we establish that this class is a two-parameter family, which 
reduces to a unique tensor $W$(up to a constant 
factor) when the complete symmetry on the four indices is required. 
We show that the superenergy density $W^{\al\ba\ga\da}u_{\al}u_{\ba}u_{\ga}u_{\da}$ relative to any timelike unit vector $u$ is positive definite and that the supermomentum density $W^{\al\ba\ga\da}u_{\ba}u_{\ga}u_{\da}$ is a timelike or a null vector 
($W^{\al\ba\ga\da}$ stands for $W$). 
Next, we find an infinite set of conserved tensors $U_{(p,q)}$ of rank $2+p+q$, that we call {\em weak superenergy tensors of order n} when $p=q=n$. We show that $U_{(1,1)}$ and $W$ yield the same total superenergy and the same total supermomentum. Then, using the canonical quantization scheme, we construct explicitly the superhamiltonian and the supermomentum operators corresponding to $W$ and to each weak superenergy tensor $U_{(n,n)}$. Finally, we exhibit a two-parameter family of superenergy tensors for an 
electromagnetic field and for a gravitational field. 
\end{abstract}

\section{Introduction}
In general relativity, the energetic content of an electromagnetic 
field propagating in a region free of charge is described by the 
well-known symmetric traceless tensor
\be
T^{\al\ba}_{\em el} = -\frac{1}{4\pi}(F^{\al\la}F^{\ba}_{\; \: \la} - 
\frac{1}{4}g^{\al\ba}F^{\rho\sa}F_{\rho\sa})\, , \lb{max}
\ee
where $F^{\al\ba}$ is the Faraday tensor. This tensor satisfies the
 conservation law \cite{foot1}
\be
{T^{\al\ba}_{\em el}}_{;\al} = 0 \lb{clm}
\ee
as a consequence of Maxwell equations with $j^{\mu}=0$. The tensor 
$T^{\al\ba}_{\em el}$ enables us to define a local density of 
electromagnetic energy as measured by an observer moving with the 
unit 4-velocity $u$ :
\be
w_{\em el}(u) = T^{\al\ba}_{\em el}u_{\al}u_{\ba} \, . \lb{dem}
\ee

It follows from (\ref{max}) that the energy density $w_{\em el}(u)$
is positive definite for any timelike unit vector $u$ :
\be
\forall u (\; g(u,u)=1 \Rightarrow  w_{\em el}(u)\geq 0\;) \lb{pos1}
\ee
and
\be
((\exists u,\; g(u, u)=1)\; w_{\em el}(u)=0\;) \Rightarrow 
T^{\al\ba}_{\em el}=0 \;(\Leftrightarrow F^{\al\ba}=0) . \lb{dfp1}
\ee

Within the framework of general relativity, however, it is well 
known that the concept of local energy density is meaningless for 
a gravitational field. An attempt to overcome this difficulty led 
to introduce the notion of superenergy tensor constructed 
with the curvature tensor $R_{\mu\nu\rho\sa}$. The first example 
of such a tensor was exhibited by Bel \cite{bel1} for a vacuum 
(see also \cite{rob1}). Then, Bel generalized his result to 
the case of an arbitrary gravitational field \cite{bel2}. By analogy 
with the tensor (\ref{max}) which may be written as
\be
T^{\al\ba}_{\em el}=-\frac{1}{8\pi}(F^{\al\la}F^{\ba}_{\; \: \la}+
\stackrel{\ast}{F}{\!}^{\al\la}\stackrel{\ast}{F}{\!}^{\ba}_{\; \la}) \, , \lb{max2}
\ee
where $\stackrel{\ast}{F}{\!}^{\al\ba}$ is the dual of $F^{\al\ba}$, the Bel tensor $B^{\al\ba\ga\da}$ is defined by
\be
2B^{\al\ba\ga\da}=R^{\al\mu\ga\nu}R^{\ba \; \da \;}_{\; \mu \; \nu}
+\ast R^{\al\mu\ga\nu} \ast R^{\ba \; \da \;}_{\; \mu \; \nu}
+{R\ast}^{\al\mu\ga\nu} {R\ast}^{\ba \; \da \;}_{\; \mu \; \nu} 
+\ast {R\ast}^{\al\mu\ga\nu} \ast {R\ast}^{\ba \; \da \;}_{\; \mu \; \nu} ,
 \lb{bel}
\ee
where $\ast$ denotes the duality operator acting on the left or on 
the right pair of indices according to its position.

The tensor defined by (\ref{bel}) has the following properties \cite{foot2}.

{\bf 1}. Its components can be expressed quadratically in terms of the curvature tensor. Indeed, it is easily checked that (\ref{bel}) 
may be rewritten in the form
\bea
B^{\al\ba\ga\da} & = & R^{\al\mu\ga\nu}R^{\ba \;  \da  \;}_{\; \mu \; \nu}
+R^{\ba\mu\ga\nu}R^{\al \; \da \;}_{\; \mu \; \nu}
-\frac{1}{2}g^{\al\ba}R^{\ga\la\mu\nu}R^{\da}_{\; \la\mu\nu} \nonumber \\
   &   & \mbox{} -\frac{1}{2}g^{\ga\da}R^{\al\la\mu\nu}R^{\ba}_{\; \la\mu\nu} 
+\frac{1}{8}g^{\al\ba}g^{\ga\da}R^{\mu\nu\rho\sa}R_{\mu\nu\rho\sa} \, . \lb{bel2}
\eea

{\bf 2}. It has the properties of symmetry
\be
B^{\al\ba\ga\da}=B^{\ba\al\ga\da}=B^{\al\ba\da\ga} \lb{sym1}
\ee 
and 
\be
B^{\al\ba\ga\da}=B^{\ga\da\al\ba} \, . \lb{sym2}
\ee

{\bf 2'}. In fact, this tensor is totally symmetric in any domain 
where $R_{\mu\nu} = 0$ :
\be
R_{\mu\nu} = 0 \; \Rightarrow \; B^{\al\ba\ga\da}=
B^{(\al\ba\ga\da)} \, .\lb{symm}
\ee 

{\bf 3}. The Bel tensor satisfies a conservation equation when the 
Einstein equations for a vacuum hold :
\be
R_{\mu\nu} = \Lambda g_{\mu\nu} \; \Rightarrow \; 
{B^{\al\ba\ga\da}}_{;\al} = 0 \, , \lb{clb}
\ee
$\Lambda$ being a cosmological constant.

{\bf 4}. For any timelike unit vector $u$, one has
\be
B^{\al\ba\ga\da}u_{\al}u_{\ba}u_{\ga}u_{\da} \; \geq \; 0 
\lb{posb}
\ee
and
\be
B^{\al\ba\ga\da}u_{\al}u_{\ba}u_{\ga}u_{\da} = 0 \; 
\Rightarrow \; B^{\al\ba\ga\da} = 0 \, . \lb{dfpb}
\ee

Properties ({\bf 3}) and ({\bf 4}) allow to regard the quantity
\be
\epsilon (u) = B^{\al\ba\ga\da}u_{\al}u_{\ba}u_{\ga}u_{\da} \,  \lb{seg}
\ee
as having the key properties of an energy density \cite{bel1,bel2}. So $\epsilon (u)$  will be called the {\em gravitational superenergy density relative to (an observer moving with) the unit 4-velocity $u$}. In the same way, one can define a {\em gravitational supermomentum density 
vector relative to u} by 
\be
p^{\al}(u) = B^{\al\ba\ga\da} u_{\ba}u_{\ga}u_{\da} \, .  \lb{smv}
\ee

This supermomentum density vector has been shown to be timelike or null for any timelike unit vector $u$ \cite{bonilla}. As a consequence, the vector 
\be
\overline{p}^{\,\mu}(u) = (\delta^{\mu}_{\al}-u^{\mu}u_{\al})B^{\al\ba\ga\da} u_{\ba}u_{\ga}u_{\da}
 \,  \lb{poy}
\ee
may be considered as playing the role of a Poynting vector for the gravitational field, as it was already proposed in \cite{bel1,bel2}.

In spite of its attractive features, the Bel superenergy tensor has not received any clear, widely accepted physical interpretation \cite{komar}${\,-\,}$\cite{mashoon}. However, the fact that $\epsilon (u)$ is positive definite plays a key role in many mathematical studies on the behavior of solutions to Einstein 
equations \cite{christodoulou, york}. It is therefore justified to pursue the study of this tensor and to examine its possible applications \cite{deser2} or extensions \cite{deser3}.
 
In particular, it is natural to ask whether it is possible to construct tensors analogous to 
(\ref{bel2}) for other fields than gravitation. This question was answered positively by Chevreton \cite{che} using the spinorial formalism on a curved spacetime. This author constructed in particular a superenergy tensor for the Maxwell field in flat spacetime and for the gravitational field within the pentadimensional Kaluza-Klein theory. Employing a different method, Komar proposed in \cite{komar} a conserved 
rank 4 tensor as a kind of Bel tensor for the Klein-Gordon field. However, the Komar tensor is less than satisfactory because it does not possess the symmetries and the positivity properties that one may expect 
for a superenergy tensor. Recently, Senovilla has found an algebraic procedure providing superenergy tensors having good properties for arbitrary fields and has given examples of such tensors for the scalar field\cite{senov1, senov2}. Nevertheless, it seems that a systematic study devoted to the Klein-Gordon field is still lacking. The main purpose of the present paper is to undertake this study within the framework of special relativity.

The plan is as follows. In Sect. 2, we give our definition of superenergy tensors of even rank $2(n+1)$ for a real scalar field satisfying the Klein-Gordon equation. Sections 3, 4 and 5 are devoted to the case where $n=1$. In Sect. 3, we establish that the rank $4$ tensors fulfilling our definition constitute a two-parameters family. In Sect. 4, we show that this  
family reduces to a unique tensor $W$ (up to an arbitrary constant factor) when 
the property of complete symmetry on the four indices is required. In Sect. 5, we show that the superenergy density is positive definite and that the supermomentum density vector is  timelike or null. In Sect. 6, we give a very simple procedure to form an infinite set of conserved tensors of rank $2+p+q$, denoted by $U_{(p,q)}$. When $p=q=n$, we find that these tensors have almost all the good properties of the superenergy tensors as defined in Sect. 2. So we call the tensor $U_{(n,n)}$ the {\em weak superenergy tensor of order $n$ }. Restricting our attention to 
the case $n=1$, we obtain the explicit relation between $U_{(1,1)}$ and one of  the super-energy tensors obtained in Sect. 4. This relation enables us to show that for 
$n=1$ the weak superenergy tensor is equivalent to the superenergy tensor
$W$ in order 
to construct  integral conserved quantities. In Sect. 7, using the 
canonical quantization procedure, we construct explicitly the superhamiltonian and the supermomentum operators associated 
with the tensor $W$ and with the weak superenergy tensors $U_{(n,n)}$. 
In Sect. 8, we indicate the two-parameter families of superenergy tensors that we have obtained with our procedure for the electromagnetic field and for the gravitational field. Finally, we gives our conclusions in Sect. 9.

\section{Superenergy tensors for a scalar field}
We consider here a real massive scalar field $\phi$ in Minkowski 
spacetime $({\cal R}^4, g)$. In what follows, the coordinates $x^{\al} = (x^{0},{\bf x})$ are always supposed to be Galilean and orthogonal, so that the metric components $g_{\al\ba}$ are 
\be
g_{\al\ba} = {\em diag}(1,-1,-1,-1) \, . \lb{mink}
\ee

We shall denote by ${\cal G}$ the inertial frame of reference corresponding to these coordinates.

We suppose that $\p$ satisfies the Klein-Gordon equation
\be
(\Box + m^2)\phi = 0 \, . \lb{kg}
\ee

This equation can be derived from the Lagrangian function 
\be
L[\phi ,\partial \phi] \equiv \frac{1}{2}g^{\al\ba}\phi_{,\al}
\phi_{,\ba} -\frac{1}{2}m^2 \phi ^2\, . \lb{lgs}
\ee

As for the electromagnetic field, it is possible to construct a symmetric, divergence-free energy-momentum tensor $T^{\al\ba}[\p]$
\be
T^{\al\ba}[\p] =\p^{,\al}\p^{,\ba} - \f \,g^{\al\ba}
(\p^{,\la}\p_{,\la} - m^{2} \p^{2}) \, . \lb{can}
\ee

It is well known that the energy density $\epsilon_{(0)}(u)= T^{\al\ba}u_{\al}u_{\ba}$ 
relative to any timelike unit vector $u$ is positive definite. It is well known also that owing to the conservation law ${T^{\al\ba}}_{,\al}
=0$, the integrals extended over the 3-space $x^{0}=const.$ 
\be
H[\p]= \int T^{00}d^{3}{\bf x'} = \int \f \, \left[(\p_{,0})^{2} +
\sum_{i}(\p_{,i})^{2} + m^{2} \p^{2} \right]d^{3}{\bf x'} \lb{en00}
\ee
and 
\be
P^{i}[\p]= \int T^{i0}d^{3}{\bf x'} = - \int \p_{,0} \p_{,i} 
d^{3}{\bf x'} \lb{mom}
\ee
can be regarded as the components in the frame ${\cal G}$ of the 
so-called energy-momentum vector $P^{\al}[\p]$. This vector is defined in any Lorentz frame by
\be
P^{\al}[\p]= \int_{\sa} T^{\al\ba}n_{\ba}d\sa \, , \lb{moment}
\ee
$\sa$ being an arbitrary spacelike 3-plane of surface element $d\sa_{\ba}
= n_{\ba}d\sa$. Since $T^{\al\ba}$ is divergence-free, the components $P^{\al}$ are constants of the motion: $H[\p]$ and $P^{i}[\p]$ respectively define the total energy and the total momentum of the field in ${\cal G}$. 

It is natural to ask whether there exist tensors $T^{\al_{1} \ba_{1}...\al_{n+1} \ba_{n+1}}$ of even rank $2(n+1) \geq 4$ which 
generalize the usual energy-momentum tensor \cite{foot3}. By analogy with the above mentioned properties of the Bel tensor, we require these 
tensors to have the following properties.

{\bf P1}. Their components $T^{\al_{1} \ba_{1}...\al_{n+1} \ba_{n+1}}$ are linear combinations of terms quadratic in $\p$ or in the derivatives 
of $\p$ of order $r \leq n+1$. More precisely, denoting by 
$\mu_{A}$ a block of indices $\{\mu_{k}, 1 \leq k \leq A \}$, we 
suppose that $T^{\al_{1} \ba_{1}...\al_{n+1} \ba_{n+1}}$ may be written in 
the form \cite{foot4}
\be
T^{\al_{1} \ba_{1}...\al_{n+1} \ba_{n+1}}= \sum^{A=n+1}_{A=0}
{C^{\al_{1} \ba_{1}...\al_{n+1} \ba_{n+1}}}_{\mu_{A} \nu_{A}} 
\p^{,\mu_{A}} \p^{,\nu_{A}} \, , \lb{quad}
\ee
where the coefficients ${C^{\al_{1} \ba_{1}...\al_{n+1} \ba_{n+1}}}_{\mu_{A}\nu_{A}}$ are tensorial quantities involving only 
the components of the metric (for $A=0$, we put $\p^{,\mu_{A}}=\p$). 

{\bf P2}. These tensors are symmetric in each pair $(\al_{i} , \ba_{i})$ of indices; they are also symmetric in the interchange of two blocks $(\al_{i} , 
\ba_{i})$ and $(\al_{j} , \ba_{j})$.

{\bf P3}. They satisfy a conservation equation when the equation of motion holds:
\be
(\Box + m^2)\phi = 0 \; \Rightarrow \; 
{T^{\al_{1} \ba_{1}...\al_{i} \ba_{i} ...\al_{n+1} \ba_{n+1}}}_{,\al_{i}} = 0 \, . \lb{csv2}
\ee

We shall call {\em superenergy tensor of order n} (or in brief {\em n-superenergy tensor}) any tensor of rank $2(n+1)$ possessing properties 
{\bf P1}-{\bf P3} \cite{foot5}. We can regard the usual energy-momentum tensor (\ref{can}) as a superenergy tensor of order $0$.

Suppose that there exists at least one $n$-superenergy tensor $T^{\al_{1} \ba_{1}...\al_{n+1} \ba_{n+1}}$. Given any timelike unit vector $u$, we can define the {\em n-supermomentum density vector} 
$\sa^{\al}_{(n)}(T, u)$ {\em relative to} $u$  by
\be
\sa^{\al}_{(n)}(T, u)= T^{\al \ba_{1}...\al_{n+1} \ba_{n+1}}u_{\ba_{1}}...u_{\al_{n+1}}
u_{\ba_{n+1}} \, . \lb{supmoment}
\ee

The {\em n-superenergy density} $\epsilon_{(n)}(T, u)$ {\em relative to} $u$ will then be defined by contracting $\sa^{\al}_{(n)}(T, u)$ with $u_{\al}$ : 
\be
\epsilon_{(n)}(T, u)= T^{\al_{1} \ba_{1}...\al_{n+1} \ba_{n+1}}u_{\al_{1}}u_{\ba_{1}}
...u_{\al_{n+1}}u_{\ba_{n+1}} \, . \lb{super}
\ee 

The tensors $T^{\al_{1} \ba_{1}...\al_{n+1} \ba_{n+1}}$ may of course be required to satisfy the additional property

{\bf P4}. The superenergy (\ref{super}) is positive definite, {\em i. e.} for each timelike unit vector $u$, one has
\be
\epsilon_{(n)}(T, u) \geq 0 \, \lb{supen}
\ee
and
\be
\epsilon_{(n)}(T, u) = 0 \; \Rightarrow \; T^{\al_{1} \ba_{1}...\al_{n+1} \ba_{n+1}} = 0 \, . \lb{defpos}
\ee

However, we shall establish in the next section that at least for $n=1$ 
property {\bf P4} restricts only the sign of the arbitrary constant factor by which one can multiply a given superenergy tensor. For this reason, it does not seem very useful to put {\bf P4} in the list 
of the fundamental requirements defining the superenergy tensors. In fact, we shall see that the most natural mean to reduce the arbitrariness on the superenergy tensors is to require these tensors to be totally symmetric. As a consequence, property {\bf P2} may be replaced by the more restrictive requirement 

{\bf P'2}. The tensor $T^{\al_{1} \ba_{1}...\al_{n+1} \ba_{n+1}}$ is 
totally symmetric.

With any superenergy tensor $T^{\al_{1} \ba_{1}...\al_{n+1} \ba_{n+1}}$ 
we can associate the tensor of rank $2n+1$ defined by
\be
\P = \int_{\sa}
T^{\al_{1} \ba_{1}...\al_{n}\ba_{n}\al_{n+1} \ba}\, n_{\ba}\, d\sa  \, . \lb{supmom}
\ee

In what follows, we assume that $\p$ and its partial derivatives of 
any order with respect to time are functions of rapid decrease at 
spatial infinity \cite{reed}. Then it results from the conservation law (\ref{csv2}) that the tensor $\P$ does not depend on the spacelike 
3-plane $\sa$. We shall call it the {\em n-supermomentum tensor 
associated with the n-superenergy tensor} $T^{\al_{1} \ba_{1}...\al_{n+1} \ba_{n+1}}$. This tensor generalizes the vector (\ref{moment}).

In the frame ${\cal G}$, we can evaluate the components $\P$  by choosing any 3-plane $x^{0}=const.$ for $\sa$. Thus we get
\be
\P = \int
T^{\al_{1} \ba_{1}...\al_{n}\ba_{n}\al_{n+1}0}(x^{0}, {\bf x'})d^{3}{\bf x'}  \, . \lb{supmom2}
\ee

Since the integrals in the r.h.s. of Eq. (\ref{supmom2}) do not depend 
on $x^{0}$, the quantities $P^{\al_{1}\ba_{1}...
\al_{n}\ba_{n}\al_{n+1}}(T)$ are constants of the motion. Among all these conserved quantities, we shall consider here only those which correspond to $\ba_{1}=\al_{2}=...= \ba_{n}=\al_{n+1}=0$. For 
the sake of brevity, let us denote by $0_{r}$ any block of $r$ timelike indices :
\be
0_{r} = \underbrace{0.....0}_{r} \, . \lb{time}
\ee

We shall call {\em (total) n-superenergy associated with the n-superenergy tensor} $T^{\al_{1} \ba_{1}...\al_{n+1} \ba_{n+1}}$ the quantity 
\be
P^{\, 0\, 0_{2n}}(T) =  \int T^{\,0\, 0_{2n+1}}(x^{0}, {\bf x'})d^{3}{\bf x'}  \, \lb{supenerg}
\ee
and {\em (total) spatial n-supermomentum associated with} $T^{\al_{1} \ba_{1}...\al_{n+1} \ba_{n+1}}$ the quantities 
\be
P^{\, i\, 0_{2n}}=  \int T^{\, i\, 0_{2n+1}}(x^{0}, {\bf x'})d^{3}{\bf x'}  
 \, . \lb{supmomt}
\ee

\section{Class of rank 4 tensors satisfying properties {\bf P1}--{\bf P3}}
The most general tensor $\T$ which fulfills conditions 
{\bf P1} and {\bf P2} may be written as \cite{foot6} 
\bea
\T & = & a\phi^{,\al\ba} \phi^{,\ga\da} + b\phi^{,\ga(\al} \phi^{,\ba)\da} 
\nonumber \\
   &   & \mbox{} + \f c_{1}(g^{\al\ba}\phi^{,\ga\la}{\phi^{\da}}_{\la} + 
g^{\ga\da}\phi^{,\al\la}{\phi^{\ba}}_{\la}) \nonumber \\
   &   & \mbox{} +c_{2}(g^{\ga(\al}\phi^{,\ba)\la}{\phi^{,\da}}_{\la}
+g^{\da(\al}\phi^{,\ba)\la}{\phi^{,\ga}}_{\la}) \nonumber \\
   &   & \mbox{} +\f d_{1}m^{2}(g^{\al\ba}\phi^{,\ga}\phi^{,\da} + 
g^{\ga\da}\phi^{,\al}\phi^{,\ba}) \nonumber \\
   &   & \mbox{} +d_{2}m^{2}(g^{\ga(\al}\phi^{,\ba)}\phi^{,\da}
+g^{\da(\al}\phi^{,\ba)}\phi^{,\ga}) \nonumber \\
   &   & \mbox{} + \f K_{1}g^{\al\ba}g^{\ga\da}
+K_{2}g^{\ga(\al}g^{\ba)\da} \, , \lb{sym} 
\eea
with 
\be
K_{s}=p_{s}\phi^{,\rho\sa}\phi_{,\rho\sa} +
q_{s}m^{2}\phi^{,\la}\phi_{,\la} + r_{s}m^{4}\phi^{2}\, , \lb{inv}
\ee
$a,b,c_{1},c_{2},d_{1},d_{2},p_{s},q_{s}$ and $r_{s}$ being dimensionless constants $(s=1,2)$.

A straightforward calculation yields:
\be
{\T}_{,\al} \equiv D^{\ba\ga\da}(\Box + m^{2})\phi + E^{\ba\ga\da}\, ,
 \lb{div1}
\ee
where $D^{\ba\ga\da}$ is the differential operator defined by
\bea
D^{\ba\ga\da} & \equiv & [a g^{\al\ba}\p^{,\ga\da} +
b g^{\al(\ga}\p^{,\da)\ba}+\f c_{1}g^{\ga\da}\p^{,\al\ba} 
 +c_{2}g^{\ba(\ga}\p^{,\da)\al}]\partial_{\al} \nonumber \\
 & & \mbox{} \hspace{20mm} + \f m^{2}[ d_{1}g^{\ga\da}\p^{,\ba}
+ 2d_{2}g^{\ba(\ga}\p^{,\da)}] \lb{div2}
\eea
and $E^{\ba\ga\da}$ is given by
\bea
E^{\ba\ga\da} & \equiv & (a+c_{2})\p^{,\al\ba}{\p^{,\ga\da}}_{\al} 
+ (b+c_{1}+c_{2}){\p_{,\al}}^{\;\ba(\ga}\p^{,\da)\al} \nonumber \\
              &        & \mbox{} +(d_{2}-a)m^{2}\p^{,\ba}\p^{,\ga\da} 
+ (d_{1}+d_{2}-b)m^{2}\p^{,\ba(\ga}\p^{,\da)} \nonumber \\ 
              &        & \mbox{} + \f g^{\ga\da}(K_{1}+\frac{c_{1}}{2}\p^{,\rho\sa}\p_{,\rho\sa}
+\frac{d_{1}-c_{1}}{2}m^{2}\p^{,\la}\p_{,\la}-\frac{d_{1}}{2}
m^{4}\p^{2})^{,\ba} \nonumber \\
              &        & \mbox{} + g^{\ba(\ga}(K_{2}+\frac{c_{2}}{2}\p^{,\rho\sa}\p_{,\rho\sa}
+\frac{d_{2}-c_{2}}{2}m^{2}\p^{,\la}\p_{,\la}-\frac{d_{2}}{2}
m^{4}\p^{2})^{,\da)} . \lb{zer}
\eea

From Eq. (\ref{zer}), it is easily seen that the conservation law
\be
{\T}_{,\al} = 0 \lb{cls}
\ee
holds for any solution to the equation of motion (\ref{kg}) if the 
invariants $K_{s}$ are defined by
\be
2K_{s}=-c_{s}\p^{,\rho\sa}\p_{,\rho\sa}+(c_{s}-d_{s})
m^{2}\p^{,\la}\p_{,\la}+ d_{s}m^{4}\p^{2} \lb{Ks}
\ee
and if the coefficients $a, b, c_{s}, d_{s}$ are solutions to the 
linear system of equations
\be
\left\{ \begin{array}{l}
a+c_{2}=0 \, , \\
b+c_{1}+c_{2}=0 \, , \\
d_{2}-a=0 \, ,  \\
d_{1}+d_{2}-b=0 \, . \\
\end{array} \right.  \lb{coef}
\ee

The above system consists in four equations for six unknown 
quantities. As a consequence, we have
\be
c_{1} = a-b \, ,\hspace{15mm} c_{2} = -a \, , 
 \lb{c}
\ee
and
\be 
d_{1} = b-a\, , \hspace{15mm} d_{2} = a \, , \lb{d} 
\ee
where $a$ and $b$ can be chosen arbitrarily.

Substituting for $c_{1}, c_{2}, d_{1}$ and $d_{2}$ from Eqs. (\ref{c}) -
(\ref{d}) into Eqs. (\ref{Ks}) and (\ref{sym}) yields the following theorem \cite{bel3}.
\begin{th}
Given a scalar field $\p$, let $\tau^{\al\ba}$ be the symmetric tensor 
defined by
\be
\tau^{\al\ba} = \p^{,\al\la}{\p^{,\ba}}_{\la} - m^{2}\p^{,\al}\p^{,\ba} 
- \f K g^{\al\ba}\, , \lb{tau}
\ee
where $K$ is the invariant
\be
K = \f \p^{,\rho\sa}\p_{,\rho\sa} - m^{2}\p^{,\la}\p_{,\la} 
+ \f m^{4}\p^{2} \, . \lb{K}
\ee

The class of rank 4 tensors fulfilling conditions {\bf P1}, {\bf P2} and {\bf P3} is the two-parameter family given by
\be
\T = a\, \T_{1} + b\, \T_{2} \, , \lb{fam}
\ee
where
\bea
\T_{1} & = & \p^{,\al\ba}\p^{,\ga\da}+ \f (g^{\al\ba}\tau^{\ga\da}
+ g^{\ga\da}\tau^{\al\ba}) \nonumber \\
       &   & \mbox{} -g^{\al(\ga}\tau^{\da)\ba}-g^{\ba(\ga}\tau^{\da)\al} 
\lb{T1}
\eea
and 
\be
\T_{2} = \p^{,\al(\ga}\p^{,\da)\ba} - \f (g^{\al\ba}\tau^{\ga\da} 
+ g^{\ga\da}\tau^{\al\ba})\, . \lb{T2}
\ee
\end{th}

A straightforward calculation allows to complete the above theorem 
by the following one.
\begin{th}
Let $u$ be an arbitrary unit timelike vector. The 1-supermomentum 
density vector $\sa^{\al}_{(1)}(T, u)$ relative to $u$ associated with the tensor $\T $ is given by
\be
\sa^{\al}_{(1)}(T,u) = (a + b)\sa^{\al}_{(1)}(T_{2}, u)\, , \lb{P}
\ee
where $\sa^{\al}_{(1)}(T_{2}, u)$ denotes the supermomentum density vector corresponding to $\T_{2}$.
\end{th}

These theorems reveal that properties {\bf P1}--{\bf P3} do not 
determine a unique  superenergy tensor (by unique, we mean of course unique up to a constant factor). Moreover, it results from 
Theorem 2 that property {\bf P4} is unable to reduce the class of 
possible tensors to a one-parameter family.

\section{Characterization of a unique superenergy tensor}  
The arbitrariness in the superenergy tensors 
is not surprising. In fact, it is easily seen that if a tensor $\T$ 
fulfills properties {\bf P1}--{\bf P4}, then the tensor 
${\widetilde{T}}^{\al \ba \ga \da}$ defined by
\be
{\widetilde{T}}^{\al \ba \ga \da} = \f (T^{\al \ga \ba \da} + 
T^{\al \da \ba \ga}) \, \lb{til}
\ee
fulfills also properties {\bf P1}--{\bf P4}. In addition, it 
is worthy to note that we have
\be
\sa^{\al}_{(1)}(\widetilde{T},u) =  \sa^{\al}_{(1)}(T,u) \, . \lb{Ptil}
\ee

Thus, $\T$ and ${\widetilde{T}}^{\al \ba \ga \da}$ have undistinguishable supermomenta density vectors.

These properties suggest that $\T _{1}$ and $\T _{2}$ are related by the 
operator $\; \widetilde{} \;$. Effectively, it is easily checked that
\be
\T _{2} = \widetilde{T}^{\al\ba\ga\da}_{1} \, . \lb{til1}
\ee

Conversely, noting that $\widetilde{\widetilde{T}} = \f (\,T + \widetilde{T})$, we get
\be
\T_{1} = 2 \widetilde{T}^{\al\ba\ga\da}_{2} - \T_{2} \, . \lb{til2}
\ee

The tensor obtained by the complete symmetrization of a tensor $\T$ 
satisfying property {\bf P2} is given by
\be
T^{(\al\ba\ga\da)} = \frac{1}{3}(\T +T^{\ba\ga\al\da}+T^{\al\ga\ba\da})\, 
, \lb{sym3}
\ee
an expression which can also be written as
\be
T^{(\al\ba\ga\da)} = \frac{1}{3}(\T +2 \tilde{T}^{\al\ba\ga\da}) \,  
\lb{sym4}
\ee
when (\ref{til}) is taken into account. From (\ref{sym4}) and (\ref{Ptil}),
 it follows that the totally symmetric tensor $T^{(\al\ba\ga\da)}$ is 
sufficient to determine the supermomentum vector density 
$\sa^{\al}_{(1)}(T,u)$ associated with $\T$.

Now, using (\ref{sym4}), (\ref{til1}) and (\ref{til2}) we get
\be
T^{(\al\ba\ga\da)}_{2} = \, T^{(\al\ba\ga\da)}_{1} 
= \frac{1}{3}(\T_{1}+2\,\T_{2}) \, . \lb{sym5}
\ee

From Theorem 1 and from Eq. (\ref{sym5}) we deduce immediately the following theorem. 

\begin{th}
Any 1-superenergy tensor $T^{\al\ba\ga\da}$ which is totally symmetric is given by
\be
T^{\al\ba\ga\da} = k \, W^{\al\ba\ga\da} \, , \lb{tot}
\ee
where $k$ is an arbitrary constant and $W^{\al\ba\ga\da}$ is defined by 
\be
W^{\al\ba\ga\da} = T^{(\al\ba\ga\da)}_{1} = T^{(\al\ba\ga\da)}_{2} \, . \lb{sym5a}
\ee

The tensor $W^{\al\ba\ga\da}$ may be written as
\bea
W^{\al\ba\ga\da} & = & \frac{1}{3}(\p^{,\al\ba}\p^{,\ga\da} +
2\p^{,\al(\ga}\p^{,\da)\ba}) \nonumber \\
                   &   & \mbox{} - \frac{1}{6}(g^{\al\ba}\tau^{\ga\da}
+ g^{\ga\da}\tau^{\al\ba} + 2g^{\al(\ga}\tau^{\da)\ba} + 
2g^{\ba(\ga}\tau^{\da)\al})\, . \lb{sym6}
\eea
\end{th}

This theorem shows that properties {\bf P1}, {\bf P'2} and {\bf P3} constitute a set of axioms allowing to select a unique 
(up to an arbitrary constant factor) 1-superenergy tensor for a scalar field.

We shall henceforth consider $W^{\al\ba\ga\da}$ as {\em the 1-superenergy 
tensor in the strict sense}. We shall put for the sake of brevity
\be
s^{\al}(u)= \sa^{\al}_{(1)}(W, u)  \lb{s1u}
\ee
and
\be
w(u)= s^{\al}(u)u_{\al} = \epsilon_{(1)}(W, u) \, , \lb{w1}
\ee
where $W$ stands for the tensor given by (\ref{sym6}). The quantities 
$s^{\al}(u)$ and $w(u)$ will be respectively regarded as ``the'' 
1-supermomentum density vector and ``the'' 1-superenergy density of the scalar field $\p$ relative to a given timelike unit vector $u$.

\section{Positivity properties of $W^{\al\ba\ga\da}$}
Introducing the projection tensor
\be
h_{\al\ba} = g_{\al\ba} - u_{\al}u_{\ba} \, , \lb{pro}
\ee
let us define the spacelike vector ${\overline{s}}^{\,\al}(u)$ by
\be
{\overline{s}}^{\,\al}(u) = h^{\al}_{\ba}s^{\ba}(u) \, . \lb{smp}
\ee

Some algebra yields for the superenergy density $w(u)$:
\bea
2\,w(u) & = & \f (u^{\mu}u^{\rho}-h^{\mu\rho})(u^{\nu}u^{\sa}-h^{\nu\sa}) \p_{,\mu\nu} \p_{,\rho\sa}  \nonumber \\
        &   & \mbox{} + m^{2}(u^{\mu}u^{\nu}-h^{\mu\nu})\p_{,\mu}\p_{,\nu} + \f m^{4} \p^{2} \lb{sen3}
\eea
and for ${\overline{s}}^{\,\al}(u)$:
\be
2\,{\overline{s}}^{\,\al}(u) =  h^{\al\rho}[(u^{\mu}u^{\sa}-h^{\mu\sa})u^{\nu}\p_{,\mu\nu}\p_{,\rho\sa}   
    +m^{2}(u^{\la}\p_{,\la})\p_{,\rho}]\, . \lb{smp2}
\ee 

It is well known that $u^{\mu}u^{\nu}-h^{\mu\nu}$ defines a positive definite metric and it is easily checked that
\be
(u^{\mu}u^{\rho}-h^{\mu\rho})(u^{\nu}u^{\sa}-h^{\nu\sa})S_{\mu\nu}
S_{\rho\sa} > 0  \lb{pos3}
\ee
for any tensor $S_{\mu\nu}$, except if $S_{\mu\nu} = 0$. Consequently, 
it follows from Eq. (\ref{sen3}) that $w(u)$ is positive definite for 
any timelike vector $u$.

Hence the following theorem.

\begin{th}
Property {\bf P4} is satisfied by the tensor $W^{\al\ba\ga\da}$.
\end{th}

Let us assume that $w(u)=0$ at a point $x$ for some $u$ such that 
$g(u,u)=1$. It results from Eq. (\ref{sen3}) that the field has the following properties at $x$ :

If $m \neq 0$, then $\p = 0, \, \p_{,\al}=0$ and $\p_{,\al\ba} = 0$.
  
If $m = 0$, then $\p_{,\al\ba} = 0$. This means that if there exists a timelike vector field $u$ such that $w(u) = 0$ everywhere, then $\p$ is necessarily of the form
\be
\p = k_{\al}x^{\al} \, , \lb{pw}
\ee
where the $k_{\al}$ are constant quantities.

We are now in a position to form in the case $n=1$ the expressions of 
the constants of the motion defined by Eqs. (\ref{supenerg}) and (\ref{supmomt}). Putting for the sake of brevity $W_{(1)}= P^{000}(W)$ and $S^{i}_{(1)}=P^{i00}(W)$, we have for the {\em 1-superenergy} of the field $\p$  
\be
W_{(1)}[\p] = \int W^{0000}(x^{0},{\bf x'})d^{3}{\bf x'} \lb{aa}
\ee
and for its {\em spatial 1-supermomentum} 
\be
S^{i}_{(1)}[\p] = \int W^{i000}(x^{0},{\bf x'})d^{3}{\bf x'} \, . \lb{ab}
\ee
The expressions of $W^{0000}$ and of $W^{i000}$ may be obtained by putting $u = \partial_{0}$ in Eqs. (\ref{sen3}) and (\ref{smp2}) respectively. We find
\bea
W^{0000}= w(\partial_{0}) = s^{0}(\partial_{0})& = & \k (\p_{,00})^{2} + \f \sum_{i} (\p_{,0i})^{2} + 
\k \sum_{i,j} (\p_{,ij})^2  \nonumber  \\
                &   & \mbox{} + \f m^{2}\left[ (\p_{,0})^{2} + 
\sum_{i}(\p_{,i})^{2} + \f m^{2}\p^{2}\right]  \lb{sen4}
\eea
and
\be
W^{i000}= \overline{s}^{\, i}(\partial_{0}) = s^{i}(\partial_{0})= -\f \left[ \p_{,00}\p_{,0i} + 
\sum_{j}\p_{,0j}\p_{,ij} + m^{2}\p_{,0}\p_{,i} \right] \, .  \lb{smp3}
\ee

We can establish an other important theorem.
\begin{th}
For any timelike unit vector $u$ the 1-supermomentum density vector 
$s^{\al}(u)$ is timelike or null : 
\be
s^{\al}(u)s_{\al}(u) \geq 0 \, . \lb{nonspa}
\ee
\end{th}

{\em Proof}. The timelike unit vector $u$ being given, let us choose the Galilean coordinates $x^{\al}$ so that $\partial_{0}=u$. From (\ref{sen4}) and (\ref{smp3}), we immediately deduce that for each $i=1,2,3$ :
\bea
s^{0}(u) \pm s^{i}(u) & = & \k \sum_{\la}(\p _{,0\la} \mp \p _{,i\la})^{2}
+ \k \sum_{\la} \sum_{j \neq i}(\p _{,\la j})^{2} \nonumber \\
                      &   & \mbox{} + \k m^{2} \left[(\p_{,0} \mp 
\p _{,i})^{2} + \sum_{j \neq i}(\p_{,j})^{2} + \sum_{\la}(\p_{,\la})^{2}
+ m^{2} \p ^{2} \right] \, . \lb{tim}
\eea

Therefore we have the three inequalities
\be
[s^{0}(u)]^{2} - [s^{i}(u)]^{2} \geq 0  \, .\lb{pos3}
\ee

Let us put $\vec{s}(u)= \overline{s}^{\,i}(u) \partial_{i}$. If $\vec{s}(u) = 0$, (\ref{pos3}) implies that $s^{2}(u) \geq 0$. So let us assume that $\vec{s}(u) \neq 0$. Then we can suppose that $\vec{s}(u)$ and $\partial_{1}$ are colinear, which implies that
\be
s^{2}(u) \equiv s^{\al}(u)s_{\al}(u) = [s^{0}(u)]^{2}-[s^{1}(u)]^{2} 
\, . \lb{tim2}
\ee
From (\ref{pos3}), we deduce that $s^{2}(u) \geq 0$. Q.E.D.

From the properties established in this section and in Sect. 4, we can conclude that $W^{\al\ba\ga\da}$ (or any tensor $k\, W^{\al\ba\ga\da}$ 
with $k > 0$) has all the good properties that one can expect for a suitable generalization of the usual energy-momentum tensor of the Klein-Gordon field.  
   
\section{Weak superenergy tensors of order $n \geq 1$}
We could of course try to determine the possible families of superenergy tensors of order $n > 1$ by using the method of Sect. 3. 
However, this procedure requires heavier and heavier calculations as $n$ is increasing. So we propose here a more simple method to construct  divergence-free tensors of rank $>2$ providing an infinite set of conserved quantities.This method
uses the property that the derivatives of any order of a solution 
to a linear field equation is still a solution. We shall here examine in detail the equivalence of the conserved quantities obtained by this 
method with the conserved superenergies and supermomenta  only in the 
case where $n=1$. The general case $n>1$ will be studied elsewhere \cite{tey}.

Let $\p_{M}$ and $\p_{N}$ be two solutions to Eq. (\ref{kg}). It 
is well known that the vector 
\be
j^{\mu}_{MN}=i(\p_{M}\p^{,\mu}_{N} - \p^{,\mu}_{M}\p_{N}) 
\lb{aj}
\ee
satisfies the conservation law
\be
{j^{\mu}_{MN}}_{,\mu} = 0 \, , \lb{ak}
\ee
a fundamental property which allows to define a scalar product on 
the space of the solutions to the Klein-Gordon equation.

By analogy with (\ref{aj}), let us put
\be
U^{\al\ba}_{MN} = \f(\p^{,\al}_{M}\p^{,\ba}_{N}+\p^{,\ba}_{M}\p^{,\al}_{N})
-\f g^{\al\ba}(\p^{,\la}_{M}\p_{N\, ,\la} - m^{2}\p_{M} \p_{N}) \, .  \lb{am}
\ee

This quantity is symmetric in $(\al ,\ba )$ and in $(M,N)$ :
\be
U^{\al\ba}_{MN} = U^{\ba\al}_{MN} = U^{\al\ba}_{NM} \, . \lb{an}
\ee
 
Calculating ${U^{\al\ba}_{MN}}_{,\al}$ we find the identity 
\be
{U^{\al\ba}_{MN}}_{,\al} \equiv \f \p^{,\ba}_{M}(\Box +m^{2})
\p_{N}+\f \p^{,\ba}_{N}(\Box +m^{2})\p_{M} \, . \lb{ap}
\ee

It is clear that for each set of solutions $\{\p_{M},\p_{N}\}$, 
$U^{\al\ba}_{MN}$ defines a divergence-free tensor symmetric 
in $(\al ,\ba)$. We note that this tensor generalizes the usual 
energy-momentum tensor (\ref{can}) since we have
\be
U^{\al\ba}_{MM} = T^{\al\ba}[\p_{M}] \, . \lb{aq}
\ee

Substituting $\p_{,\mu_{1}\mu_{2} ...\mu_{p}}$ for $\p_{M}$ and  
$\p_{,\nu_{1}\nu_{2} ...\nu_{q}}$ for $\p_{N}$ into Eq. (\ref{am}) we 
obtain the following quantities :
\bea
{U^{\al\ba}}_{\mu_{1} ...\mu_{p}\nu_{1} ...\nu_{q}} & = & 
{\p^{(,\al}}_{\mu_{1}...\mu_{p}}{\p^{,\ba)}}_{\nu_{1}...\nu_{q}}-\f g^{\al\ba}({\p^{,\la}}_{\mu_{1}...\mu_{p}}\p_{,\la \nu_{1}...\nu_{q}} \nonumber \\
                                                    &   & \mbox{} -m^{2}
\p_{,\mu_{1}...\mu_{p}}\p_{,\nu_{1}...\nu_{q}}) \, , \lb{ar}
\eea
which constitute the components of a tensor of rank $(2+p+q)$ that we shall denote by the intrinsic notation $U_{(p,q)}$.

The tensor $U_{(p,q)}$ presents very interesting features. It satisfies property {\bf P1}. It is symmetric in $(\al,\ba)$. It is also completely symmetric 
in $(\mu_{1} ,...,\mu_{p})$ and in $(\nu_{1} ,... ,\nu_{q})$. 
Moreover, it satisfies the conservation law
\be
{U^{\al\ba}}_{\mu_{1} ...\mu_{p}\nu_{1} ...\nu_{q},\al}=0 \,  \lb{au}
\ee
as a consequence of Eq. (\ref{ap}).

Let us now restrict our attention to the case where $p=q=n$. It is easily seen that for any timelike unit vector $u$
\be
{U^{\al\ba}}_{\mu_{1} ...\mu_{n}\nu_{1} ...\nu_{n}}
u_{\al}u_{\ba}u^{\mu_{1}}...u^{\mu_{n}}u^{\nu_{1}}...u^{\nu_{n}}\geq 0 \, . \lb{av}
\ee

Consequently, the tensor $U_{(n,n)}$ has almost all the good 
properties of a superenergy tensor of order $n$. For this reason, we call $U_{(n,n)}$ the {\em weak superenergy tensor of order n} or the {\em weak n-superenergy tensor}. In conformity with Eq. (\ref{aq}) we consider that $U_{(0,0)}$ is the usual energy-momentum tensor (\ref{can}).

In the inertial frame ${\cal G}$, we can define the {\em weak n-superenergy} $U_{(n)}[\p]$ by 
\be
U_{(n)}[\p] = \int {U^{00}}_{0_{n}0_{n}}(x^{0},{\bf x'}) 
d^{3}{\bf x'}  \lb{ava}
\ee
and the {\em weak spatial n-supermomentum} $R^{i}_{(n)}[\p]$ by
\be
R^{i}_{(n)}[\p]= \int {U^{i0}}_{0_{n}0_{n}}(x^{0},{\bf x'})d^{3}{\bf x'}
 \, , \lb{avb}
\ee
where we have used the notation defined by (\ref{time}). These quantities 
may be explicitly written as
\be
U_{(n)}[\p] = \f \int \left[(\p_{,0_{n+1}})^{2} + 
\sum_{i}(\p_{,i 0_{n}})^{2} + m^{2} (\p_{,0_{n}})^{2}
\right]d^{3}{\bf x'} \, \lb{ave1}
\ee
and
\be
R^{i}_{(n)}[\p] = - \int \p_{,0_{n+1}}\p_{,i 0_{n}}
d^{3}{\bf x'} \, . \lb{avf1}
\ee

It follows from (\ref{au}) that these integral quantities are constants 
of the motion.
 
What is the relation between the tensors $U_{(n,n)}$ and the superenergy 
tensors defined in Sect. 2 ? We shall treat here this problem  only in the case where $n=1$. It follows from (\ref{ar}) that the contravariant components of $U_{(1,1)}$ are  
\bea
U^{\al\ba\mu\nu} & = & \f (\p ^{,\al\mu}\p ^{,\ba\nu} +
\p ^{,\ba\mu}\p ^{,\al\nu}) \nonumber \\
                  &   & \mbox{} -\f g^{\al\ba} (\p ^{,\la\mu}
{\p ^{,\nu}}_{\la} - m^{2} \p ^{,\mu} \p ^{,\nu} ) \, . \lb{aw}
\eea
Comparing with (\ref{T2}),it is easily seen that
\be
T^{\al\ba\mu\nu}_{2} =  U^{\al\ba\mu\nu}-\f (\tau^{\al\ba} - \f K g^{\al\ba})
g^{\mu\nu} \, . \lb{ax}
\ee

A straightforward calculation yields
\be
\tau^{\al\ba} - \f K g^{\al\ba} = {U^{\al\ba\la}}_{\la} - 
m^{2} T^{\al\ba} \, , \lb{ay}
\ee
where ${U^{\al\ba\la}}_{\la} = g_{\mu\nu} U^{\al\ba\mu\nu}$. As a consequence, $T^{\al\ba\mu\nu}_{2}$ is given by
\be
T^{\al\ba\mu\nu}_{2} =  U^{\al\ba\mu\nu}-
\f ({U^{\al\ba\la}}_{\la} - m^{2} T^{\al\ba})g^{\mu\nu} \, . \lb{az}
\ee

It is easy to check that $U^{\al\ba\mu\nu}-\f ({U^{\al\ba\la}}_{\la} - m^{2} T^{\al\ba}) g^{\mu\nu}$ is symmetric in the interchange of the blocks $(\al,\ba)$ and $(\mu,\nu)$ of indices. Consequently, 
$T^{\al\ba\mu\nu}_{2}$ may be written in a form which renders more explicit all its symmetries :
\bea
2T^{\al\ba\mu\nu}_{2} & = & U^{\al\ba\mu\nu}+ U^{\mu\nu\al\ba} - 
\f ({U^{\al\ba\la}}_{\la} - m^{2} T^{\al\ba}) g^{\mu\nu} \nonumber \\
                     &   & \mbox{} - \f ({U^{\mu\nu\la}}_{\la} - m^{2} T^{\mu\nu}) g^{\al\ba}  \, . \lb{ba}
\eea

We are now in a position to state an important theorem about the constants of the motion of order one.
\begin{th}
The 1-superenergy $W_{(1)}[\p]$ and the spatial 1-supermomentum $S_{(1)}^{i}[\p]$ are respectively given by 
\be
W_{(1)}[\p]=U_{(1)}[\p]+[Surf] \lb{bh}
\ee
and
\be
S^{i}_{(1)}[\p] = R^{i}_{(1)}[\p]+[Surf] \, , \lb{bi}
\ee
where $[Surf]$ denotes surface terms which cancel if $\p $ and its derivative $\p_{,0}$ are functions of rapid decrease at spatial 
infinity.
\end{th}

{\em Proof}. It results from (\ref{sym5a}) and (\ref{sym3}) that 
$W^{\al 000}=T^{\al 000}_{2}$. Therefore we deduce from (\ref{az}) that
\be
W^{\al 000}=U^{\al 000}- \f ({U^{\al 0\la}}_{\la} - m^{2}T^{\al 0}) \, . 
\lb{bia}
\ee

When $\p$ satisfies the Klein-Gordon equation, each term ${U^{\al 0\la}}_{\la} - m^{2}T^{\al 0}$ is a $3$-divergence. Indeed, it is easily 
seen that in this case
\be
{U^{00\la}}_{\la} - m^{2}T^{00} = - \{\f \p_{,i}[\p^{,ij} - 
\da^{ij}(\triangle \p - 2m^{2} \p)]\}_{,j}  \lb{bd}
\ee
and
\be
{U^{i0\la}}_{\la} - m^{2}T^{i0} = \{ \p_{,0}[\p^{,ij} - 
\da^{ij}(\triangle \p - m^{2} \p)]\}_{,j} \, , \lb{bg}
\ee
where $\triangle$ is the usual Laplacian operator on ${\cal R}^{3}$. Thus 
the theorem is established. Q. E. D.

\section{The superhamiltonian and the supermomentum operators in 
quantum field theory}
The results obtained in Sect. 5 and 6 enable us to construct an 
infinite set of superhamiltonian and supermomentum operators in 
quantum field theory.  

Within the canonical quantization procedure, a real solution $\p$ to 
the Klein-Gordon equation becomes a Hermitian operator which can be 
expanded on the basis of the plane wave solutions as
\be
\p ({\bf x}, t) = \frac{1}{(2\pi)^{3/2}}
\int d^{3}{\bf k}\frac{1}{\sqrt{2\om}}\left [a({\bf k})e^{-ikx} + 
a^{\dag}({\bf k})e^{ikx} \right ]_{k^{0}= \omega_{k}} \, , 
\lb{fop}
\ee
with
\be
\om = \sqrt{{\bf k}^{2} + m^{2}} \, , \lb{pul}
\ee
the operators $a({\bf k})$ and $a^{\dag}({\bf k})$ their Hermitian 
conjugates satisfying the commutation relations
\be
[a({\bf k}), a^{\dag}({\bf k'})] = \da ^{(3)}({\bf k}-{\bf k'}) \, , 
\lb{com1}
\ee
\be 
[a({\bf k}), a({\bf k'})]=[a^{\dag}({\bf k}), a^{\dag}({\bf k'})] = 0 
\, . \lb{com2}
\ee

$W_{(1)}$ and $S^{i}_{(1)}$ become what we shall call the {\em 1-superhamiltonian} and the 
{\em spatial 1-supermomentum operators}, respectively. Substituting 
$\p$ from (\ref{fop}) into (\ref{sen4}) and (\ref{smp3}), and 
integrating over the space yield
\be
{\widehat{W}}_{(1)} = \frac{1}{2}\int d^{3}{\bf k} \,\omega^{3}_{k}
[a^{\dag}({\bf k})a({\bf k}) + a({\bf k})a^{\dag}({\bf k})] \, , \lb{shm}
\ee
\be
{\widehat{S}}^{i}_{(1)} =  \int d^{3}{\bf k} \, k^{i} \omega^{2}_{k} \, a^{\dag}({\bf k})a({\bf k}) \, . \lb{smop}
\ee

These expressions are remarkably simple and may be compared with 
those of the usual Hamiltonian and momentum component operators 
deduced respectively from (\ref{en00}) and from (\ref{mom})
\be
\widehat{H} = \f \int d^{3}{\bf k}\, \om \,  [a^{\dag}({\bf k})a({\bf k}) 
+ a({\bf k})a^{\dag}({\bf k})] \, , \lb{ham}
\ee
\be
{\widehat{P}}^{i} = \int d^{3}{\bf k} \, k^{i}\,  a^{\dag}({\bf k})a({\bf k}) 
\, . \lb{mom2}
\ee

More generally, the {\em weak n-superhamiltonian} corresponding to (\ref{ave1}) is given by
\be
{\widehat{U}}_{(n)} = \f \int d^{3}{\bf k}\, \om^{2n+1} \, [a^{\dag}({\bf k})a({\bf k}) + a({\bf k})a^{\dag}({\bf k})] \,  \lb{hamn}
\ee
and the {\em weak spatial n-supermomentum} deduced from (\ref{avf1}) is
\be
{\widehat{R}}^{i}_{(n)} = \int d^{3}{\bf k} \,  k^{i}\om^{2n} \, a^{\dag}({\bf k})a({\bf k}) 
  \, . \lb{momn}
\ee

A comparison of Eqs. (\ref{hamn}) and (\ref{momn}) in the case of $n=1$ 
with Eqs. (\ref{shm}) and (\ref{smop}) shows that
\be
{\widehat{W}}_{(1)} = {\widehat{U}}_{(1)} \, \lb{ham11}
\ee
and that
\be
{\widehat{S}}^{i}_{(1)} = {\widehat{R}}^{i}_{(1)} \, , \lb{mom11}
\ee
equations which can also be immediately deduced from (\ref{bh}) and 
(\ref{bi}).

\section{Extension to other fields}
The systematic procedures developed here can be extended to 
electromagnetism and to other linear theories for higher spin fields. For the electromagnetic field treated in flat spacetime, {\em e. g.}, the transposition of our definitions and reasonings is obvious. In particular, it can be checked that the class of {\em divergence-free} rank 4 superenergy tensors built from $F^{\al\ba}$ is a two-parameter family : one of the generators is the tensor $M^{\al\ba\ga\da}$ obtained by Chevreton\cite{che}, the other one is the tensor ${\widetilde{M}}^{\al\ba\ga\da}$, where $\; \widetilde{} \;$ is 
the operation defined by Eq. (\ref{til}).

In fact, our method can also be applied to the gravitational field. 
Again, it is found that the class of tensors 
$T_{gr}^{\al\ba\ga\da}$ possessing properties {\bf 1}, {\bf 2} and 
{\bf 3} enumerated in Introduction is a two-parameter family given by 
\be
T_{gr}^{\al\ba\ga\da}= a_{1}\,B^{\al\ba\ga\da}+ 
a_{2}\,{\widetilde{B}}^{\al\ba\ga\da}\, , \lb{bk} 
\ee
where $a_{1}$ and $a_{2}$ are arbitrary constants. From (\ref{bel2}) 
and (\ref{til}), it results that ${\widetilde{B}}^{\al\ba\ga\da}$ may 
be  written as
\bea
{\widetilde{B}}^{\al\ba\ga\da} & = & R^{\al\mu\ba\nu}R^{(\ga \;  \da)  \;}_{\; \; \, \mu \; \nu}
 +R^{\mu\al\nu(\ga}R^{\da )\;  \ba  \;}_{\; \mu \; \; \nu} \nonumber \\
   &   & \mbox{} -R^{\la\mu\nu (\al}g^{\ba )(\ga}R^{\da )}_{\;\nu\mu\la}
    +\frac{1}{8}g^{\al(\ga}g^{\da)\ba}R^{\mu\nu\rho\sa}
R_{\mu\nu\rho\sa} \, . \lb{bl}
\eea

Using the symmetries of the Riemann tensor, it is easily checked that 
for any timelike vector $u$
\be
{\widetilde{B}}^{\al\ba\ga\da}u_{\al}u_{\ba}u_{\ga}u_{\da}=
 B^{\al\ba\ga\da}u_{\al}u_{\ba}u_{\ga}u_{\da} \, . \lb{bm}
\ee

Consequently, ${\widetilde{B}}^{\al\ba\ga\da}$ satisfies property {\bf 4}, which implies that $T_{gr}^{\al\ba\ga\da}$ is positive definite if and only if $\, a_{1}+a_{2}>0$.

Since we have ${\widetilde{B}}^{(\al\ba\ga\da)} = B^{(\al\ba\ga\da)}$, we deduce from (\ref{bk}) that the totally symmetric gravitational superenergy tensors of rank 4 are proportional to 
\bea
B^{(\al\ba\ga\da)} & = & 2R^{\mu(\al\ba}_{\; \; \; \; \; \; \; \nu}
{R_{\mu }}^{\ga\da )\nu}-g^{(\al\ba}R^{\ga}_{\; \la\mu\nu}
R^{\da )\la\mu\nu} \nonumber \\
  &   & \mbox{} + \frac{1}{8}g^{(\al\ba}g^{\ga\da)}R^{\mu\nu\rho\sa}
R_{\mu\nu\rho\sa} \, . \lb{bp}
\eea

With Eq. (\ref{bp}), we recover a tensor recently obtained  by Robinson \cite{rob2}.  

\section{Conclusions}
We have proposed a definition of the superenergy tensors of even rank $2(n+1)$ for the Klein-Gordon field. We have determined the entire class of rank 4 tensors $
T^{\al\ba\ga\da}$ obeying our definition. Clearly, the tensors $T^{\al\ba\ga\da}$ 
constitute good generalizations of the usual energy-momentum tensor. They form a 
two-parameter family. This last feature is not embarrassing, however, because the unicity (up to an arbitrary factor) can be obtained by requiring the total symmetry in the four indices.

The fact that the tensor $T^{\al\ba\ga\da}_{2}$ given by Eq. (\ref{T2}) is also obtained 
by Senovilla raises the problem of whether our definition is equivalent to the one 
proposed in \cite{senov1} and \cite{senov2}. This problem is open. 

We would like to emphasize the interest of the linear operator 
$\;\widetilde{}\;$  which transforms a rank 4 superenergy tensor into an other one of the same rank (see Sect. 4). This operator works in any spacetime. Its existence explains why the set 
of rank 4 superenergy tensors cannot reduce to a one-parameter family if the total symmetry is not required.
 
For the Klein-Gordon field, we have formed an infinite set of divergence-free tensors $U_{(n,n)}$ of rank $2(n+1)$ which have almost all the good properties of the superenergy tensors. The tensors 
$U_{(n,n)}$ yield a class of constants of the motion which constitute an 
acceptable generalization of the total energy and of the total momentum of the field. The problem of finding the relation between these tensors and the superenergy tensors of rank $2(n+1)$ is solved here for $n=1$, but remains open for $n>1$.

We shall add that tensors similar to $U_{(n,n)}$ can also be constructed for the electromagnetic field. This question is under current investigation. 

Finally, let us emphasize that the systematic procedure developed in 
Sect. 3 can be applied without difficulty to the electromagnetic field and to the gravitational field. We have briefly indicated the main results that we have obtained in this way.
 
\section*{Acknowledgements}
We are deeply indebted to Ll. Bel for initiating our interest in the problem as well as 
for many stimulating discussions and  encouragements. We are also grateful to J. M. M. Senovilla and to J.-Cl. Houard for very useful comments and for hepful suggestions.

\end{document}